\documentclass[10pt,journal,letterpaper]{IEEEtran}
\usepackage{graphicx}
\usepackage{amsmath}
\usepackage{cite}
\usepackage{xcolor}
\newcommand{\ignore}[1]{}

\ifCLASSINFOpdf
\else
\fi

\begin{document}
%
\title{A Probabilistic Behavioral RTL Simulation Framework for Probabilistic Compute Fabrics}
%
%
%

\author{Zehuan Wang,
        Chaohui Li,
        Ruihong Yin,
        Xunqi Li,
        and Chris H. Kim%
\thanks{This work was supported in part by the
Defense Advanced Research Projects Agency (DARPA)
Quantum-Inspired Classical Computing (QuICC) program
under Air Force Research Laboratory (AFRL)
contract FA8750-22-C-1034, and by the
National Science Foundation (NSF) under
Grants 2230963 and 2142248.
(Corresponding author: Chris H. Kim.)}%
\thanks{All authors are with the Department of Electrical
and Computer Engineering, University of Minnesota,
Minneapolis, MN 55455 USA
(e-mail: chriskim@umn.edu).}%
}

\maketitle

\begin{abstract}
Coupled oscillator Ising hardware can produce different solution qualities
across repeated runs of the same problem, while a deterministic behavioral
model used for the design yields a single outcome under fixed conditions. Capturing this stochastic variability during design time is essential for evaluating how closely simulation reflects
silicon behavior. This work presents a probabilistic behavioral RTL
simulation framework that introduces Gaussian timing variation into coupling delays. This variation changes
oscillator timing during evolution, producing different outcomes across
runs while retaining the existing chip control and readout blocks.
The framework is evaluated against measurements from two chips: a 28-nm
CMOS chip with 66 spins and a 16-nm FinFET chip with 49 spins, using 250
problems run 100 times repetitively. 
Compared with deterministic simulation, the model reduces the average
absolute gap between sorted simulation and silicon outcomes by 14.0\%
and 28.4\% on the two chips, respectively. These results demonstrate that
a simple behavioral timing variation model can be a useful tool for designing and verifying probabilistic computing chips.
\end{abstract}


%

\section{Introduction}

\IEEEPARstart{C}{oupled} oscillator Ising machines provide a hardware approach to combinatorial optimization. Optimization problems can be mapped to an Ising model, where spin configurations represent candidate solutions and the Hamiltonian defines their energy~\cite{lucas2014}. These machines encode spins in oscillator phases and use interactions among oscillators to evolve toward low-energy configurations~\cite{mallick2020}. CMOS implementations include locally coupled arrays~\cite{moy2022}, capacitively coupled oscillator networks~\cite{liu2024tcasii}, all-to-all connected chips~\cite{lo2023,cilasun2025}, and systems with digital search support~\cite{li2025esser,yin2026fpgaasic}. Repeated runs on physical oscillator networks can produce different solution qualities, making run-to-run variability an important consideration in the chip verification flow.

Register-transfer level (RTL) simulation in Verilog or VHDL is a standard design methodology for digital chips, but it can also be extended to simulate the behavior of coupled-oscillator Ising chips. The basic idea is to export the behavioral model of small analog or time-based building blocks and then simulate the entire network of couplers and oscillators to confirm that the oscillator phases settle to a near-ground state. One limitation of existing behavioral simulation approaches is that they are deterministic; i.e., they do not account for the stochastic behavior of oscillators governed by noise, manufacturing variation, temperature changes, voltage fluctuations, and other non-idealities found in real silicon.

Prior work has compared transistor-characterized event-driven simulation with silicon solution distributions~\cite{kumar2025droid}. The goal of these simulation models is to achieve SPICE-level accuracy with faster simulation time, rather than serving as part of an industry-standard verification flow that enables full-chip verification alongside other digital and mixed-signal IP blocks. Other studies have examined the effects of random initial phases and oscillator frequency mismatch on solution outcomes~\cite{graber2024firefly}, and incorporated stochastic phase fluctuations into analog oscillator models~\cite{chou2019weighted}.
To model run-to-run variability, this work introduces Gaussian timing perturbations into coupler delays at each transition event. We compare the resulting simulation outcomes with measurements from Ising chips fabricated in 28-nm and 16-nm CMOS, and examine how the variation magnitude affects solution quality and agreement with silicon.

\ignore{
The main contributions of this work are:
\begin{itemize}
    \item \textbf{Probabilistic RTL timing model:} We introduce Gaussian
    timing variation into pull-up and pull-down delays at each oscillator transition in the existing behavioral RTL model. This allows repeated simulations of the same problem to
    produce different outcomes while preserving the original chip control and readout flow.

    \item \textbf{Timing variation analysis:} We study how the
    variation magnitude affects solution quality through repeated
    simulations. The results show that larger timing variation does not always lead to better solutions, indicating that the variation magnitude must be selected carefully.

    \item \textbf{Silicon validation:} We compare deterministic
    and probabilistic simulations with measurements from 28-nm
    and 16-nm chips.
\end{itemize}
}

\section{Background}

\subsection{Ising Model}

Many combinatorial optimization problems can be mapped to an Ising
model~\cite{lucas2014}, where the energy of a system with $N$ spins
$s_i\in\{-1,+1\}$ is given by
\begin{equation}
H = - \sum_{i<j} J_{ij}s_i s_j - \sum_i h_i s_i,
\label{eq:ising}
\end{equation}
where $J_{ij}$ is the coupling strength between spins $i$ and $j$,
and $h_i$ is the local field of spin $i$. The optimization objective is encoded in
$J_{ij}$ and $h_i$, and minimizing (\ref{eq:ising}) gives an optimal
solution to the mapped problem. An Ising chip searches for low-energy
spin configurations. In this work, solution quality is measured by
the Ising energy, with lower energy indicating a better solution
for the same problem.

\subsection{Coupled-Oscillator Ising Machines}
\label{sec:oim}

In a coupled ring-oscillator Ising chip, each spin is
represented by a ring oscillator, with its value encoded
in the oscillator's phase~\cite{moy2022,lo2023}.
Coupling between oscillators $i$ and $j$ changes their
transition timing according to the sign of $J_{ij}$
and their relative phase. This favors in-phase or
anti-phase locking, driving the system toward
lower-energy configurations~\cite{mallick2020,moy2022}.
Subharmonic injection locking (SHIL) favors two stable
phase states separated by $180^\circ$, allowing the
oscillator phases to represent binary
spins~\cite{lo2023,cilasun2025}.
The final spin configuration depends on the timing of
oscillator transitions, and small changes in these delays
can lead to different solution outcomes.
Section~\ref{sec:method} describes how these timing
changes are modeled.

\section{Simulation and Silicon Testing}
\label{sec:method}

\subsection{System Overview}

\begin{figure}[t]
    \centering

    \textbf{(a) Silicon chip comparison}\\[3pt]

    {\footnotesize
    \setlength{\tabcolsep}{2.5pt}
    \renewcommand{\arraystretch}{1.15}

    \begin{tabular}{@{}lcc@{}}
        \hline
        \textbf{Specification} &
        \textbf{28-nm CMOS} &
        \textbf{16-nm FinFET} \\
        \hline

        Technology &
        28-nm HPC+ CMOS &
        16-nm FinFET \\

        Connectivity &
        All-to-all &
        All-to-all \\

        Coupling matrix &
        $70\times70$ &
        $50\times50$ \\

        Problem spins &
        66 &
        49 \\

        Die photo &
        \raisebox{-0.5\height}{
            \includegraphics[
                width=0.24\columnwidth
            ]{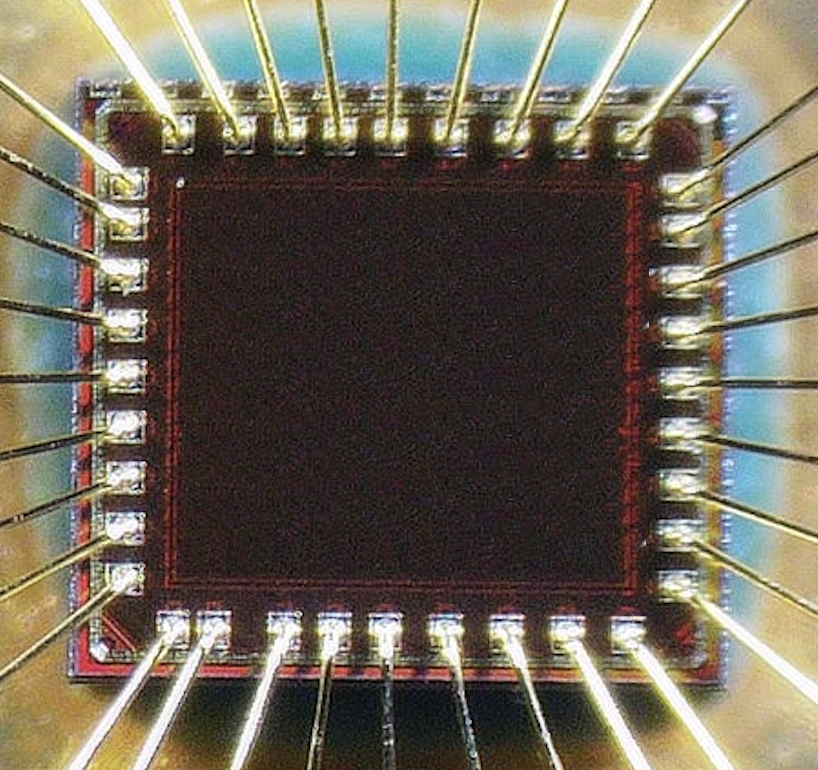}
        } &
        \IfFileExists{COBIFINFET_diephoto.png}{
            \raisebox{-0.5\height}{
                \includegraphics[
                    width=0.24\columnwidth
                ]{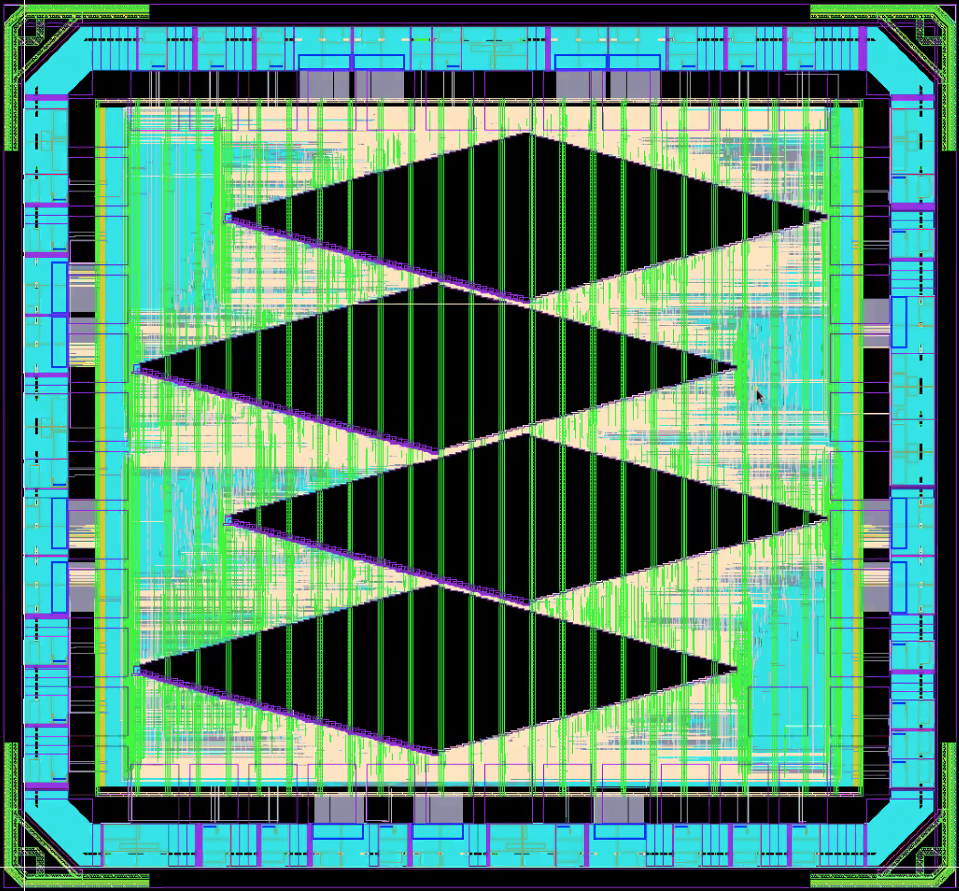}
            }
        }{
            \raisebox{-0.5\height}{
                \fbox{
                    \parbox[c][0.70cm][c]{
                        0.20\columnwidth
                    }{\centering Die photo\\TBD}
                }
            }
        } \\

        Die size &
        $1.10\times1.00$ mm$^2$ &
        $1.05\times0.97$ mm$^2$ \\

        Core $V_{\mathrm{DD}}$ &
        0.9 V &
        0.8 V \\

        I/O $V_{\mathrm{DD}}$ &
        1.8 V &
        1.8 V \\

        System clock &
        769 MHz (sim.) &
        667 MHz (sim.) \\

        Integer weights &
        $[-14,+14]$ &
        $[-14,+14]$ \\

        Majority vote &
        4-of-7 &
        4-of-7 \\

        \hline
    \end{tabular}
    }

    \vspace{5pt}

\textbf{(b) 28-nm chip architecture and spin mapping}\\[2pt]

\includegraphics[
    width=0.96\linewidth
]{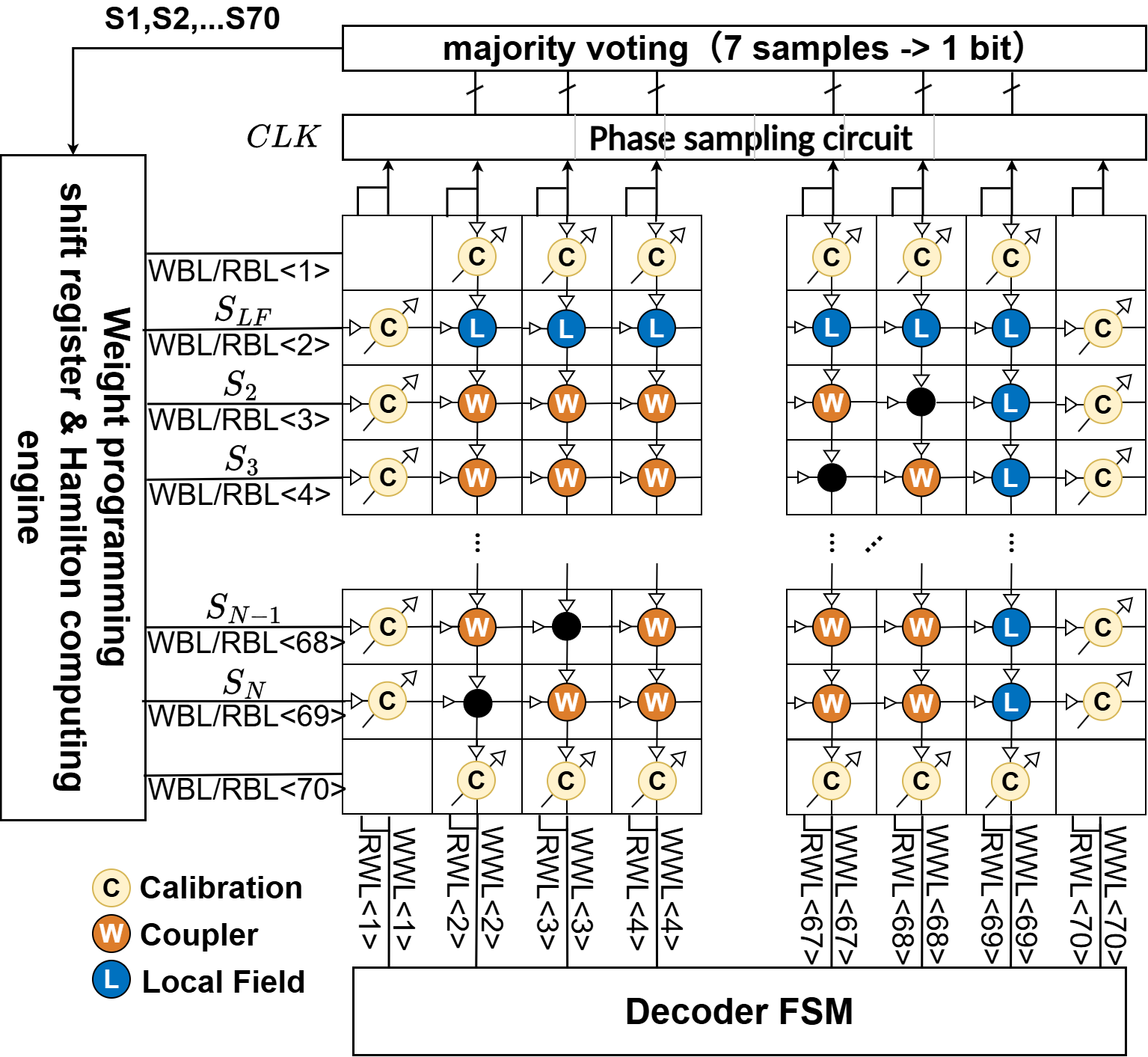}

\caption{Evaluated silicon test chips.
(a) Specifications of the 28-nm CMOS and 16-nm FinFET chips.
(b) Architecture of the 28-nm Ising solver, including
programmable coupling cells (W), local-field cells (L),
calibration cells (C), digital control, and majority-vote
readout. Of the 70 oscillator indices, 66 represent
problem spins.}

\label{fig:system_overview}
\end{figure}

Fig.~\ref{fig:system_overview}(a) compares the two silicon
implementations evaluated in this work. Both chips employ
programmable all-to-all coupled ring-oscillator arrays.
The 28-nm CMOS core uses a $70\times70$ coupling matrix
with 66 problem spins, while the 16-nm FinFET core uses
a $50\times50$ matrix with 49 problem spins.

Fig.~\ref{fig:system_overview}(b) illustrates the architecture
of the 28-nm chip. Each spin is represented by the phase
of a ring oscillator. Programmable coupling cells (W)
store 4-bit signed weights in $[-7,+7]$ and implement
pairwise interactions between oscillators. Local-field
cells (L) support individual spin biases, while calibration
cells (C) provide auxiliary circuit functions. The shift
register and decoder FSM configure and control the array.
Of the 70 oscillator indices, 0 and 69 are reserved boundary
indices, while 35 and 36 are associated with the central
subharmonic injection-locking (SHIL) structure. The
remaining 66 indices represent problem spins.

Each oscillator is sampled
seven times, and a 4-of-7 majority vote determines its
binary output. The four reserved indices are excluded
to obtain the 66-spin solution. Although the chip includes
a digital Hamiltonian engine for local-search refinement
and energy evaluation, the 28-nm experiments in this work
bypass digital spin refinement and evaluate the original
oscillator solutions. Our proposed timing perturbations
are applied to the coupler delays during oscillator
evolution, while preserving the chip control and readout.

\begin{figure*}[!t]
    \centering
    \includegraphics[width=0.95\textwidth]{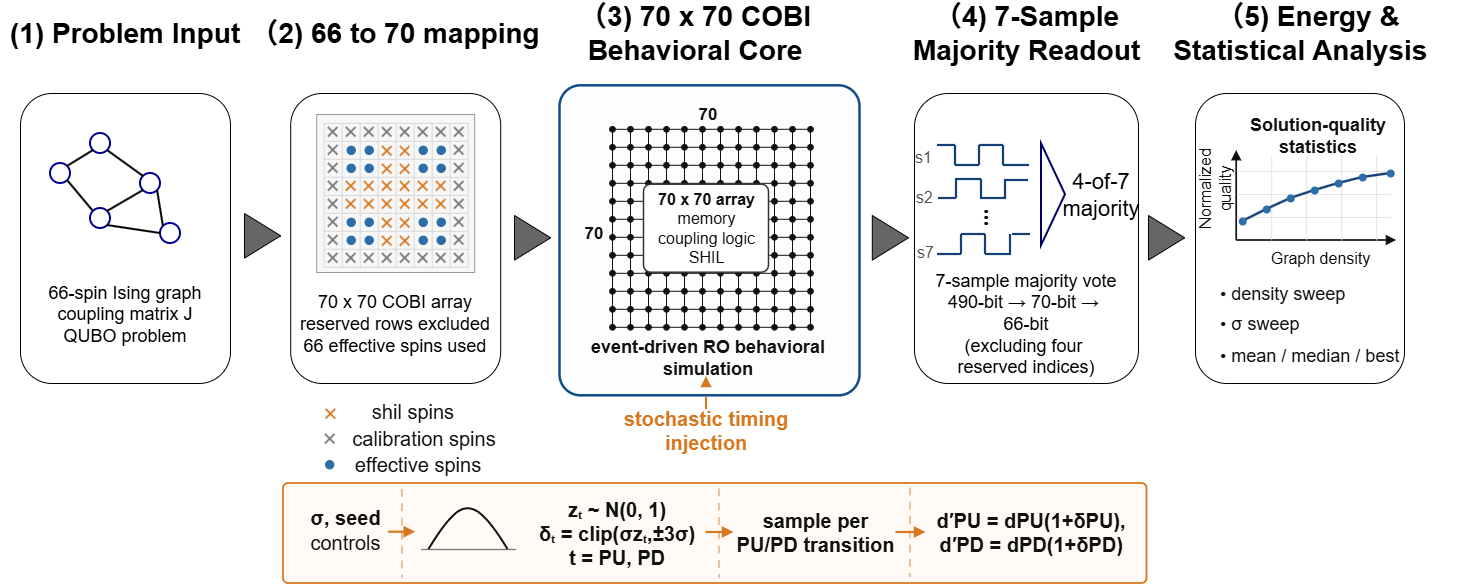}
    \caption{Behavioral simulation and silicon-comparison flow, shown for
    the 28-nm core (66 problem spins, $70\times70$ coupling matrix).
    The 16-nm core has 49 problem spins and a $50\times50$ matrix.
    Both use event-wise PU/PD timing perturbation and seven-sample
    majority-vote readout.}
    \label{fig:framework}
\end{figure*}

\begin{figure}[!htbp]
    \centering
    \includegraphics[width=0.60\columnwidth]{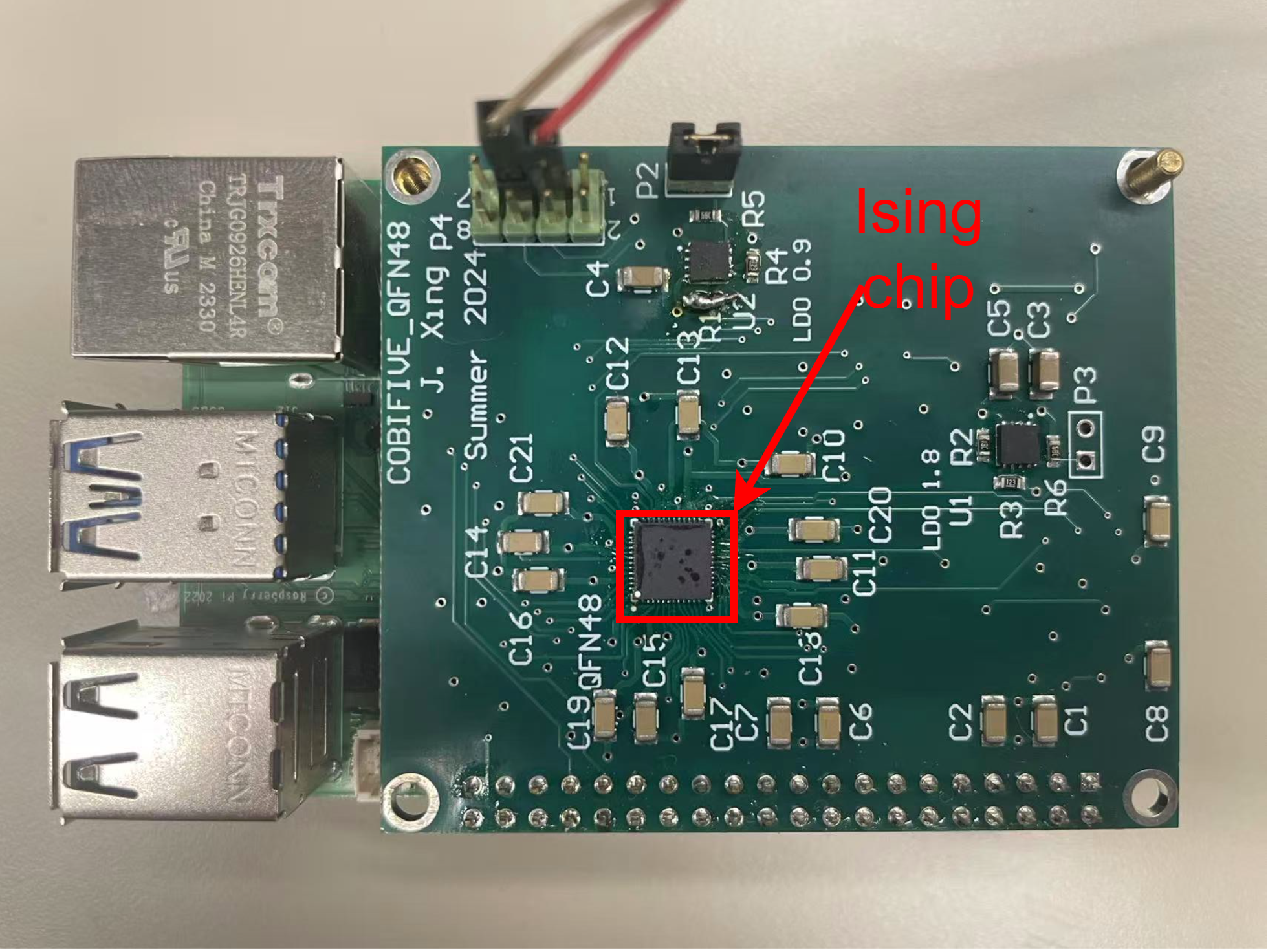}
    \caption{Test board mounted on a Raspberry Pi host for silicon measurements.}
    \label{fig:rpi}
\end{figure}

\subsection{Probabilistic Simulation Framework}

Fig.~\ref{fig:framework} shows the simulation and evaluation
flow for the 28-nm coupled oscillator Ising core. Random graph problems are converted
into hardware-compatible QUBO/Ising inputs. Logical weights
in $[-7,7]$ are mapped symmetrically to positions $(i,j)$
and $(j,i)$, giving effective coupling weights in $[-14,+14]$. This corresponds to about a 5 bit weight resolution which is on-par with commercial quantum annealers and sufficient for most hardware based Ising applications. 
Graph density is defined as the fraction of nonzero logical
couplings. The 16-nm core is a scaled version of the 28-nm core with minor changes such as fewer number of oscillators and updates to the control RTL.

The full-chip Verilog behavioral model includes the
controller/FSM, coupled-oscillator core, and sampling/readout
logic. The pull-up (PU) and pull-down (PD) delays correspond
to low-to-high and high-to-low output transitions of each coupler unit, respectively.
Each core retains its original coupling-dependent nominal delays. For repeated runs of the same problem, the programmed couplings remain fixed, while event-wise timing perturbations
vary with the random seed.

Timing perturbations are applied during the oscillator
phase evolution rather than to the final Ising energy.
Changes in the relative timing of transitions can
affect subsequent events and lead to different final
spin configurations.

After evolution, seven D-flip-flops (DFFs) sample each
oscillator using time-interleaved clocks derived from the
reference oscillator signal. A 4-of-7 majority vote
determines the binary spin value $\{-1,+1\}$.
For the 28-nm experiments, digital spin refinement is
bypassed, and the four reserved indices are excluded
to obtain the 66-spin oscillator solution.
The corresponding Ising energy is then evaluated
without local-search refinement and normalized using
the best-known solution from QBSolv~\cite{booth2017qbsolv}.

Fig.~\ref{fig:rpi} shows the PCB used for silicon measurements of both 28-nm and 16-nm test chips.


\begin{figure*}[!t]
    \centering

    \begin{minipage}[t]{0.47\textwidth}
        \centering
        \textbf{(a) 28-nm CMOS}\\[-1pt]
        \includegraphics[width=\linewidth]{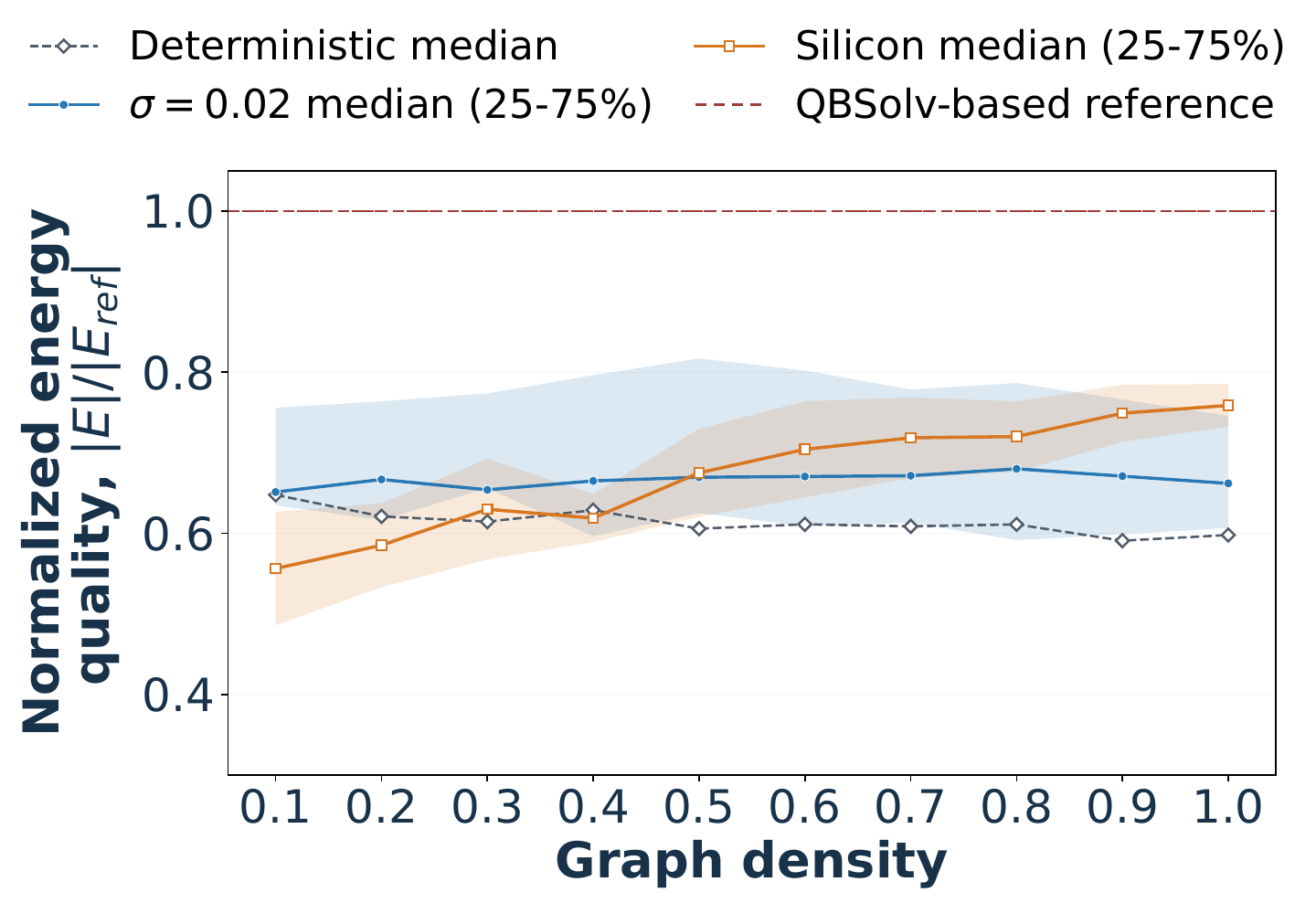}
    \end{minipage}
    \hfill
    \begin{minipage}[t]{0.47\textwidth}
        \centering
        \textbf{(b) 16-nm FinFET}\\[-1pt]
        \includegraphics[width=\linewidth]{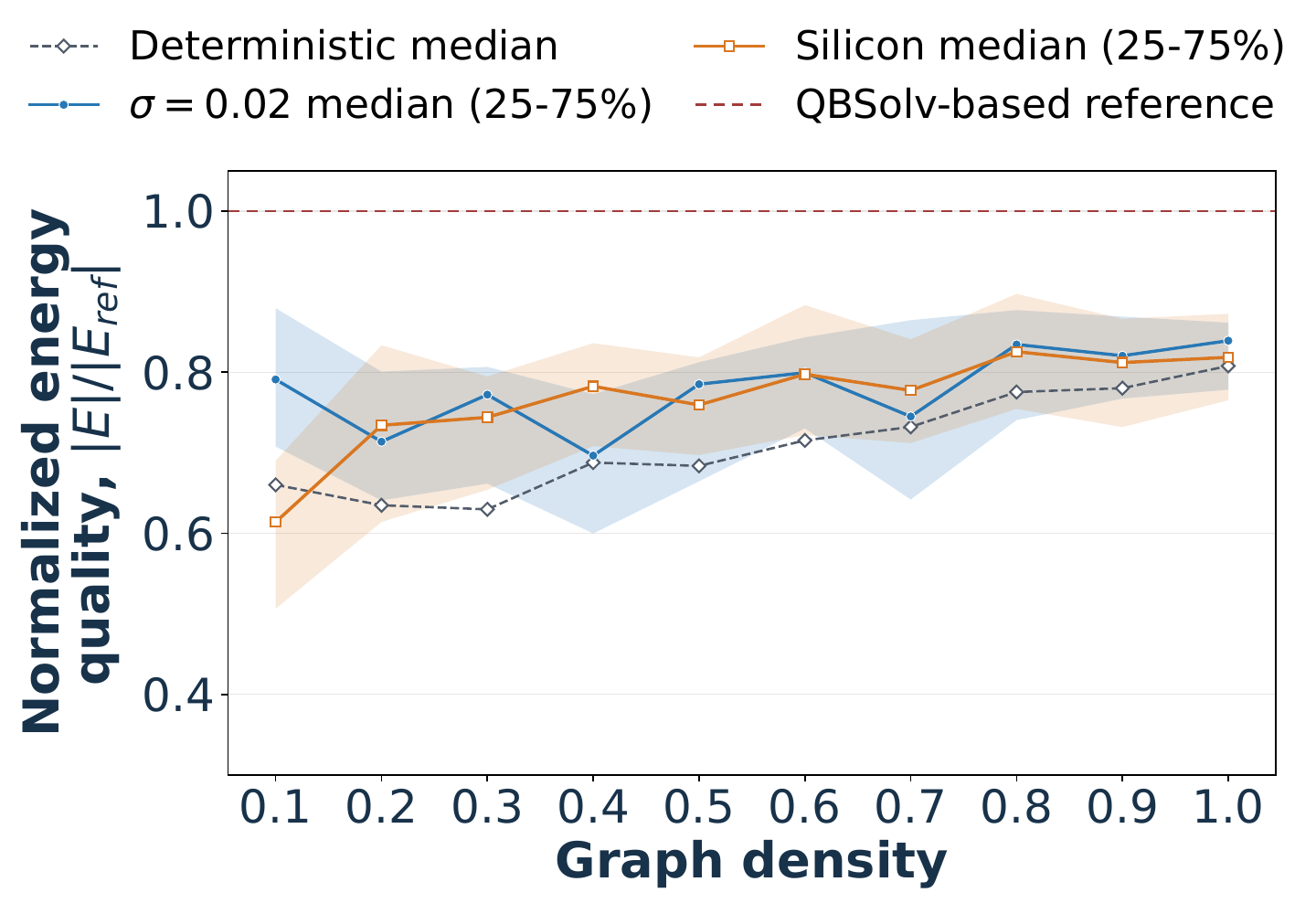}
    \end{minipage}
    \caption{Silicon data versus simulated solution quality for
    (a) 28-nm CMOS (66 problem spins) and (b) 16-nm FinFET (49 problem
    spins). Curves show medians across 25 problems per density; shading
    shows the 25th--75th percentile range. Probabilistic
    curves use $\sigma=0.02$ for both chips; $Q=1$ corresponds to the QBSolv reference energy.}
    \label{fig:quality_density}
\end{figure*}
\subsection{Timing-Variation Instantiation}

To model timing noise in silicon, we introduce Gaussian
delay perturbations into the nominal PU and PD delays
of the coupler cells. For each transition, a random
relative delay perturbation is generated from a Gaussian
distribution. The parameter $\sigma$ represents the
standard deviation of this distribution before clipping,
expressed as a fraction of the nominal delay.
The perturbation is clipped to $[-3\sigma,3\sigma]$: $z \sim \mathcal{N}(0,1)$, where $\delta = \min(3\sigma, \max(-3\sigma, \sigma z))$. The perturbed delay is given by $d' = d_{\mathrm{nom}}(1+\delta)$, where $d_{\mathrm{nom}}$ is the nominal PU or PD delay determined by the coupling-dependent timing rules in the original behavioral model. A new Gaussian sample is generated for each transition.
The implementation is shown in the following
Verilog-style pseudocode.

\begin{center}
\begin{minipage}{0.96\columnwidth}
\footnotesize
\hrule
\vspace{3pt}
\textbf{Verilog-Style Event-Wise Pull Up / Pull Down Delay Perturbation}
\vspace{3pt}

\begin{verbatim}
always @(transition_event) begin
  d_base = (transition_direction == PU) ?
    nominal_PU_delay(coupling_state) :
    nominal_PD_delay(coupling_state);
  next_state = (transition_direction == PU);
  z = (SIGMA == 0.0) ? 0.0 :
      gaussian_sample(transition_direction);
  delta = clip(SIGMA*z, -3*SIGMA, 3*SIGMA);
  #(d_base*(1.0+delta)) output_state <= next_state;
end
\end{verbatim}

\vspace{2pt}
\hrule
\end{minipage}
\end{center}

Setting $\sigma=0$ recovers the deterministic model.
We selected $\sigma$ using 300 28-nm graph instances
separate from the 250 validation instances in
Fig.~\ref{fig:quality_density}(a).
For each graph, we tested
$\sigma\in\{0.01,0.02,0.03,0.05\}$ and identified
the value whose simulated median Ising energy was
closest to the silicon median. The closest setting
was $\sigma=0.02$ for 202 graphs (67.3\%),
$0.03$ for 87 (29.0\%), and $0.01$ for 11
(3.7\%); none selected $0.05$.
We therefore fixed $\sigma=0.02$ for both chips,
without retuning for 16 nm. The selection targets
agreement with silicon rather than the lowest
simulated energy. Here, $\sigma=0.02$ corresponds
to a 2\% relative-delay standard deviation before
clipping, not a measured process-variation level.

\subsection{Experimental Setup}

We evaluate simulation-to-silicon agreement using separate problem sets for the 28-nm and 16-nm chips. For each chip, 25 independent Ising/QUBO problems are
generated at each graph density from 0.1 to 1.0
in steps of 0.1, resulting in 250 problems per chip. Each problem is defined by its coupling matrix. The 28-nm and 16-nm cores use 66 and 49 problem spins, respectively.

For each problem, deterministic simulation produces one
result, while probabilistic simulation at $\sigma=0.02$
is run 100 times and silicon is measured over 100 runs.
All three use the same programmed problem and Ising
energy evaluation.

For both chips, digital refinement is bypassed,
so the reported energies correspond to the original
majority-voted oscillator solutions. This allows
the effect of timing perturbations to be evaluated
directly without additional changes to the spin
configurations from digital refinement.
The 28-nm and 16-nm RTL models retain their respective
nominal PU/PD delays, with only the perturbation
magnitude shared between the two implementations.

\begin{figure}[!htbp]
    \centering
    \includegraphics[width=\linewidth]{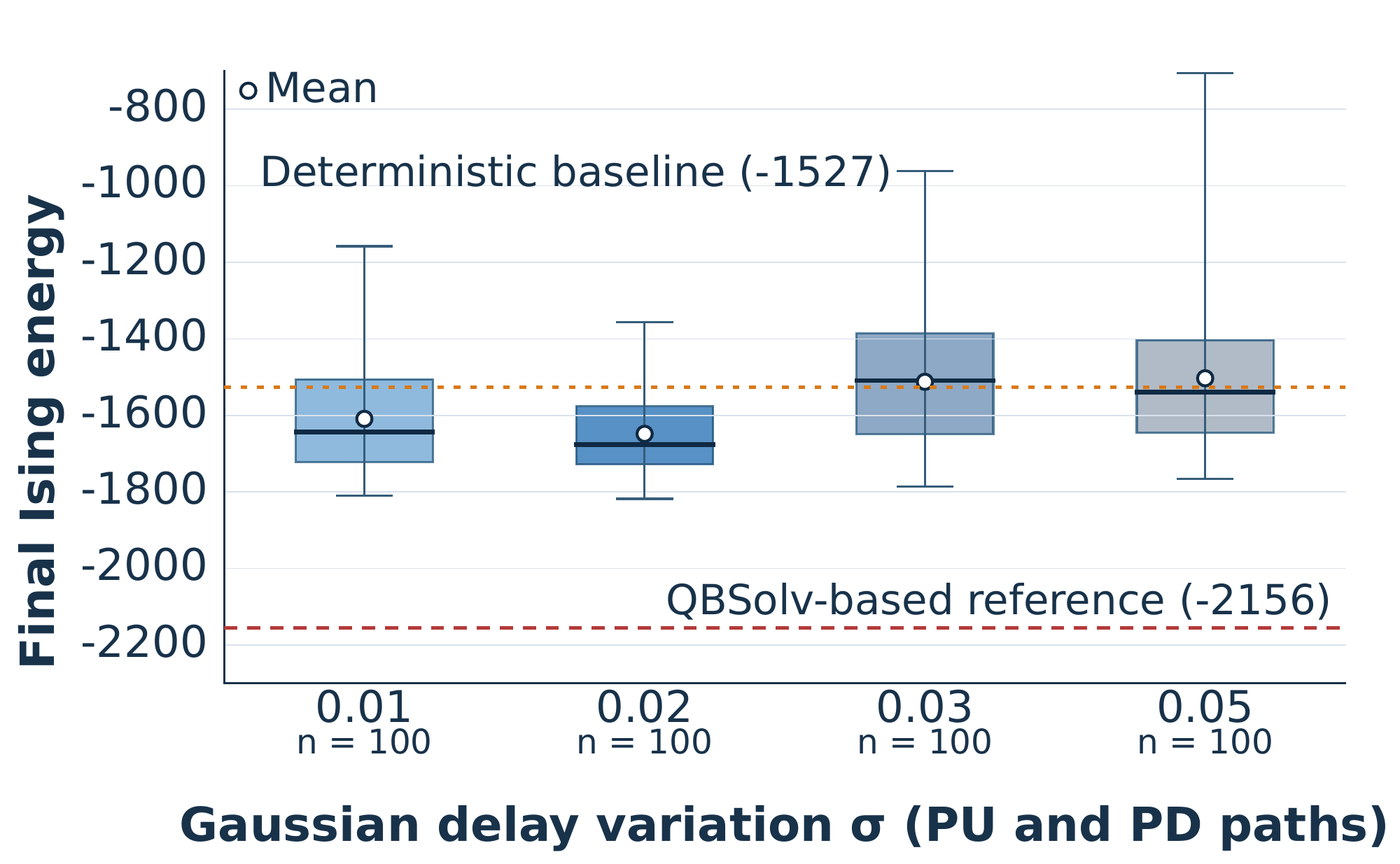}

\caption{Final Ising energy versus Gaussian delay
perturbation for a fixed Ising problem at graph
density 0.5 using the 28-nm RTL model
(100 runs per $\sigma$).
Boxes mark the 25th--75th percentile range;
bars are medians, whiskers are extrema, and
circles are means. The deterministic and QBSolv
software solver reference energies are
$-1,527$ and $-2,156$, respectively;
lower energy is better.}
    \label{fig:sigma_sweep}
\end{figure}

The fixed-instance sweep in Fig.~\ref{fig:sigma_sweep} uses a separate
problem to examine $\sigma$ sensitivity. For delay-scale context, we
simulate the schematic-level 28-nm coupler cell in SPICE and foundry models at TT, 0.9~V, 25~$^\circ$C. We also run 100 Monte Carlo samples with local mismatch at the TT global corner.

Let $E_{\mathrm{ref},d,i}$ denote the best known software solution obtained from QBSolv~\cite{booth2017qbsolv} for instance $i$ at density $d$. For run $r$ of source $s$, normalized
solution quality is
\begin{equation}
Q_{d,i,s}^{(r)}
=
\frac{\left|E_{d,i,s}^{(r)}\right|}
     {\left|E_{\mathrm{ref},d,i}\right|}.
\end{equation}
Here $s$ is the simulation or silicon source; deterministic simulation
uses the same normalization. Note that for the negative Ising energies studied, larger
$Q$ means lower energy and hence a better solution.

For each problem, we take the median $Q$ over 100 runs of silicon or
probabilistic simulation; deterministic simulation contributes one
value. In Fig.~\ref{fig:quality_density}, plotted medians and
25th--75th percentile ranges are across the 25 problems at each density,
not across runs of a single problem.

We quantify the simulation-to-silicon difference by averaging
absolute gaps in representative normalized quality:
\begin{equation}
G_s=\frac{1}{M}\sum_{j=1}^{M}
\left|\widetilde{Q}_{s,j}-
\widetilde{Q}_{\mathrm{sil},j}\right|,
\label{eq:average_gap}
\end{equation}

where $s$ denotes deterministic or probabilistic simulation,
and $\widetilde{Q}_{s,j}$ and
$\widetilde{Q}_{\mathrm{sil},j}$ denote the median normalized
solution qualities from simulation and silicon for
problem $j$, respectively. For deterministic simulation,
the single result is used instead of a median.
The absolute differences are averaged across all 250 problems ($M=250$) for each chip. A smaller $G_s$ indicates closer simulation-to-silicon agreement.

For each problem, we separately sort the 100 normalized
solution quality values from simulation and silicon.
We calculate the absolute difference between the two
values at each corresponding position and average
these differences over all 100 positions.
The resulting value is then averaged across 250 problems.
For deterministic simulation, its single result for
each problem is repeated 100 times before comparison.

For Fig.~\ref{fig:sigma_sweep}, one fixed problem is simulated with
$\sigma\in\{0.01,0.02,0.03,0.05\}$ using seeds 0--99 for each setting
(400 runs total); $\sigma=0$ is the deterministic baseline. The Hamiltonian energies of the deterministic baseline and QBSolv software solver are $-1,527$ and $-2,156$, respectively.
The improvement fraction is the proportion of 100 runs with
$E^{(r)}<E_{\mathrm{det}}$, where lower energy is better.
We report mean, median, standard deviation, quartiles, and extrema. The improvement fractions describe a single random Ising problem, not all graphs.

\section{Comparison Results and Discussion}

\subsection{Cell-Level Timing Characterization}

At nominal process corner, the coupler circuit delays are 10.2~ps (low-to-high) and 10.7~ps (high-to-low). In 100 local-mismatch Monte Carlo samples,
the low-to-high delay is $10.2\pm0.37$~ps (3.66\% relative standard
deviation); the high-to-low delay varies by 4.31\%. This few-percent
spread provides context for the $\sigma$ sweep but does not calibrate
$\sigma$, since transistor mismatch is static (i.e. does not change with time) while our model samples variation per event.
The nominal PU and PD delay characteristics are
retained separately in the RTL model, while the
same relative perturbation magnitude is applied
to both transition directions.

\subsection{Comparison With Silicon Measurements}

For the 28-nm CMOS chip, the probabilistic model at
$\sigma=0.02$ reduces the average per-problem
median-quality gap to silicon from 0.0853 to 0.0548
(35.7\%), as shown in Fig.~\ref{fig:quality_density}(a).
The average gap between sorted simulation and silicon
solution distributions also decreases by 14.0\%.
Compared with deterministic simulation, the
probabilistic model achieves closer agreement at
6 of 10 graph densities, although the simulated
density trend remains flatter than that measured
in silicon.

For the 16-nm FinFET chip, the same $\sigma=0.02$
reduces the median-quality gap from 0.065 to 0.041
(37.0\%) and the sorted-distribution gap by 28.4\%,
as shown in Fig.~\ref{fig:quality_density}(b).
Closer agreement is observed at 8 of 10 graph
densities. Although the two chips retain different
nominal PU/PD delay rules, the same perturbation
setting improves simulation-to-silicon agreement
without retuning for the 16-nm model.

\subsection{Sensitivity to Timing-Perturbation Magnitude}

Fig.~\ref{fig:sigma_sweep} compares the final Ising
energies for a fixed problem using the 28-nm RTL
model at a graph density of 0.5. The same coupling
matrix is used across all perturbation settings,
with 100 runs per setting. The effect of $\sigma$
is non-monotonic. At $\sigma=0.02$, the mean and
median energies are $-1648.2$ and $-1676.5$,
respectively, with the lowest standard deviation
among the tested settings. In addition, 81\% of
runs outperform the deterministic baseline.

Increasing $\sigma$, however, does not consistently
improve solution quality. At $\sigma=0.05$, 53\%
of runs still achieve lower energy than the
deterministic baseline, yet the mean energy is
worse. The improvement fraction only indicates
how often a run outperforms the baseline, whereas
the mean energy also depends on the magnitude
of the energy differences. Thus, more frequent
improvements do not necessarily lead to better
average solution quality.

\section{Conclusion}
We present a probabilistic behavioral RTL framework for the functional verification of coupled-oscillator-based Ising chips, validated against 28-nm CMOS and 16-nm FinFET silicon. Rather than sacrificing simulation throughput for SPICE-level transistor detail, our methodology efficiently verifies probabilistic compute fabrics while accurately mirroring physical silicon behavior—giving designers first-silicon confidence. To capture the intrinsic run-to-run variability missed by deterministic simulation, the framework injects event-wise Gaussian timing perturbations. Applying a uniform noise level ($\sigma = 0.02$) across both chips reduces the average absolute error between simulated and silicon median solution qualities by 35.7\% (28-nm) and 37.0\% (16-nm) across 250 benchmark problems. Furthermore, when comparing distribution profiles over 100 runs per problem, the discrepancy drops by 14.0\% and 28.4\%, respectively.


%



\begin{thebibliography}{10}

\bibitem{lucas2014}
A. Lucas, ``Ising formulations of many NP problems,''
\emph{Front. Phys.}, vol. 2, Art. no. 5, 2014.

\bibitem{mallick2020}
A. Mallick, M. K. Bashar, D. S. Truesdell, B. H. Calhoun,
S. Joshi, and N. Shukla, ``Using synchronized oscillators to
compute the maximum independent set,'' \emph{Nat. Commun.},
vol. 11, Art. no. 4689, 2020.

\bibitem{moy2022}
W. Moy, I. Ahmed, P.-W. Chiu, J. Moy, S. S. Sapatnekar, and
C. H. Kim, ``A 1,968-node coupled ring oscillator circuit for
combinatorial optimization problem solving,'' \emph{Nat. Electron.},
vol. 5, pp. 310--317, 2022.

\bibitem{liu2024tcasii}
Y. Liu \emph{et al.}, ``A 1024-spin scalable Ising machine with
capacitive coupling and progressive annealing method for combination
optimization problems,'' \emph{IEEE Trans. Circuits Syst. II, Exp. Briefs},
vol. 71, no. 12, pp. 5009--5013, 2024.

\bibitem{lo2023}
H. Lo, W. Moy, H. Yu, S. S. Sapatnekar, and C. H. Kim,
``An Ising solver chip based on coupled ring oscillators with a
48-node all-to-all connected array architecture,''
\emph{Nat. Electron.}, vol. 6, pp. 771--778, 2023.

\bibitem{cilasun2025}
H. C{\i}lasun \emph{et al.}, ``A coupled-oscillator-based Ising chip
for combinatorial optimization,'' \emph{Nat. Electron.}, vol. 8,
pp. 537--546, 2025.

\bibitem{li2025esser}
C. Li \emph{et al.}, ``A coupled oscillator based Ising chip with a
fully-connected annealing core, a Hamiltonian engine, and a RISC-V
gradient search for solving 2,500+ spin combinatorial optimization
problems,'' in \emph{Proc. IEEE ESSERC}, 2025, pp. 297--300.

\bibitem{yin2026fpgaasic}
R. Yin \emph{et al.}, ``An FPGA-ASIC co-design framework for
capacity-constrained physics-based Ising chips,'' in \emph{Proc.
IEEE/ACM ICCAD}, 2026, doi: 10.1145/3831252.3834138.

\bibitem{kumar2025droid}
A. Kumar, R. S., C. H. Kim, U. R. Karpuzcu, and S. S. Sapatnekar,
``DROID: Discrete-time simulation for ring-oscillator-based Ising
design,'' \emph{Sci. Rep.}, vol. 15, Art. no. 18643, 2025.

\bibitem{graber2024firefly}
M. Graber and K. Hofmann, ``Firefly: A versatile experimental platform
for oscillator-based Ising machines,'' \emph{IEEE Trans. Circuits Syst. I,
Reg. Papers}, vol. 71, no. 12, pp. 5944--5955, 2024.

\bibitem{chou2019weighted}
J. Chou, S. Bramhavar, S. Ghosh, and W. Herzog,
``Analog coupled oscillator based weighted Ising machine,''
\emph{Sci. Rep.}, vol. 9, Art. no. 14786, 2019.

\bibitem{booth2017qbsolv}
M. Booth, S. P. Reinhardt, and A. Roy, ``Partitioning optimization
problems for hybrid classical/quantum execution,'' D-Wave Systems,
Burnaby, BC, Canada, Tech. Rep. 14-1006A-A, 2017.


\end{thebibliography}
\end{document}